\documentclass[prl,twocolumn,amsmath,amssymb,superscriptaddress]{revtex4-2}
\usepackage{graphicx,amsmath,relsize,epstopdf,color,mathtools,bm,newtxtext,newtxmath,braket,rotating}
\usepackage[hyphenbreaks]{breakurl}
\usepackage[colorlinks=true,linkcolor=blue,citecolor=blue,urlcolor =blue]{hyperref}
\usepackage[normalem]{ulem}
\usepackage[table,xcdraw]{xcolor}

\newcommand{\Tr}{\mathop{\mathrm{Tr}} \nolimits}

\usepackage{soul,xcolor}
\begin{document}
\setstcolor{red}

\title{Intrinsic Sensitivity Limits for Multiparameter Quantum Metrology}

\author{Aaron Z. Goldberg}
\affiliation{National Research Council of Canada, 100 Sussex Drive, Ottawa, Ontario K1A 0R6, Canada}
\affiliation{Department of Physics, University of Toronto, 60 St. George Street, Toronto, Ontario, M5S 1A7, Canada}
%\email{aaron.goldberg@nrc-cnrc.gc.ca}

\author{Luis~L. S\'{a}nchez-Soto}
\affiliation{Departamento de \'{O}ptica, Facultad de F\'{\i}sica, Universidad Complutense, 28040 Madrid, Spain}
\affiliation{Max-Planck-Institute f\"{u}r die Physik des Lichts, 91058 Erlangen, Germany}

\author{Hugo Ferretti}
\affiliation{Department of Physics, University of Toronto, 60 St. George Street, Toronto, Ontario, M5S 1A7, Canada}

\begin{abstract}
The quantum Cram\'er-Rao bound is a cornerstone of modern quantum metrology, as it provides the ultimate precision in parameter estimation. In the multiparameter scenario, this bound becomes a matrix inequality, which can be cast to a scalar form with a properly chosen weight matrix. Multiparameter estimation thus elicits tradeoffs in the precision with which each parameter can be estimated. We show that, if the information is encoded in a unitary transformation, we can naturally choose the weight matrix as the metric tensor linked to the geometry of the underlying algebra $\mathfrak{su}(n)$, with applications in numerous fields. This  ensures an intrinsic bound that is independent of the choice of parametrization. 
\end{abstract}

\maketitle

\emph{Introduction.---} 
A central challenge in quantum metrology is to develop strategies for enhancing the precision of parameter estimation.  The quantum Fisher information (QFI), and the associated quantum Cram\'er-Rao bound (QCRB), are invaluable tools for this task~\cite{Helstrom:1976ij,Holevo:2003fv}, as they characterize the ultimate precision attainable for different classes of probe states. This is crucial to identifying quantum resources that lead to an enhancement in sensitivity versus classical strategies. 

It is well established that the maximal sensitivities achievable for the estimation of a single parameter using particle-separable and arbitrary quantum probe states are the shot noise and Heisenberg limits, respectively~\cite{Paris:2009aa}. However, much less is known about the corresponding  bounds for the simultaneous estimation of multiple parameters.  Multiparameter quantum metrology finds many important applications in fields as diverse  as imaging~\cite{Genovese:2016vd,Ang:2016aa,Rehacek:2017ab}, field sensing~\cite{Tsang:2011tr,Koschorreck:2011ub,Baumgratz:2016aa}, sensor networks~\cite{Sidhu:2017vd,Nokkala:2018tj,Rubio:2020uq}, and remote sensing~\cite{,Ansari:2021vh} to cite but a few examples.  In this case, the QCRB is a matrix inequality and the ultimate bound is generally not saturable for all parameters.  This is because the corresponding optimal observables may be incompatible~\cite{Zhu:2015wt,Heinosaari:2016wb,Ragy:2016aa} and one cannot simultaneously achieve the optimal precision for each individual parameter. Several comprehensive reviews highlight recent progress in this subject area~\cite{Szczykulska:2016aa,Sidhu:2020aa,Albarelli:2020aa,Polino:2020aa,Rafal:2020aa,Goldberg:2021uw}.

These difficulties have fueled the search for tighter bounds, which prove to be handy for practical implementations. The Holevo Cram\'er-Rao bound (HCRB)~\cite{Holevo:2003fv} epitomizes the conundrums associated with incompatible observables: its main tenet is to map the matrix QCRB onto a \textit{scalar} inequality by using a positive-definite weight matrix and then optimize this scalar bound over all physically viable measurement procedures for a given probe state. In this manner, one obtains a weighted mean square error that has to be minimized. This is considered hard to evaluate, even numerically, because it is defined through a constrained minimization over a set of operators. Closed results are known only for very simple models~\cite{Suzuki:2016tt,Matsumoto:2002tu,Bradshaw:2017vx} and the numerical tractability of calculating the HCRB for finite-dimensional systems has been recently considered~\cite{Albarelli:2019to,Sidhu:2021tp}.

Surprisingly, little attention has been paid to the properties of the weight matrix: its only role is to give relative weights to the different parameter estimate variances, so different choices lead to different optimal probe states and experimental designs. This entails a significant ambiguity in real-world problems that inherently involve numerous parameters. One idea is to always choose the weight matrix that gives the worst possible total error for an optimal measurement strategy~\cite{Belliardo:2021we}. In this Letter, we discuss whether that matrix can be chosen in a \emph{natural} way.

To provide a proper answer, we first assume that the parameters are imprinted onto the probe state via a unitary transformation; i.e., we assume that the parameters belong to the group SU($n$). From a geometrical viewpoint, the associated Lie algebra $\mathfrak{su} (n)$ can be endowed with a metric tensor, the Killing-Cartan form, with all the desirable properties~\cite{Helgason:1978vb}. When this metric is used as our weight matrix, we obtain a QCRB with intrinsic properties, independent of the choice of parametrization. Such a QCRB can significantly facilitate finding the corresponding optimal states {for estimating all of the parameters with these intrinsic weights}. We demonstrate the power of this bound for the broad scenario of estimating all of the parameters of an SU($n$) operation, which generalizes known multiparameter estimation results for rotation~\cite{Baumgratz:2016aa} and multiphase~\cite{Humphreys:2013} estimation problems. Because of the ubiquity of unitary transformations, this paves the way for solving new problems such as determining the limits of three-dimensional (3D) polarimetry~\cite{Ellis:2005un,Gil:2014wg,Gil:2017uj}, measuring systems of qudits~\cite{Thew:2002tc,Molina:2004dz,Lima:2011wt,Rambach:2021wt}, characterizing multiport interferometers~\cite{Reck:1994aa,Tan:2013ti,Clements:2016aa,Harris:2016aa}, which have found recent applications in boson sampling \cite{,Aaronson:2011tj,Tillmann:2013ux,Crespi:2013aa}, and beyond.

\emph{Setting the scenario.---}
We are interested in estimating multiple parameters simultaneously. The typical scheme requires some $d$-dimensional vector of parameters $\bm{\theta}= (\theta_{1}, \ldots, \theta_{d})^{\top} \in \mathbb{R}^{d}$ to be imprinted on a (pure) probe state $\ket{\psi}$, which is shifted by applying a corresponding unitary transformation ${U} (\bm{\theta}) \in \mathrm{SU}(n)$ that encodes the parameters $\bm{\theta}$ (the superscript $\top$ denotes the transpose). The output state $\ket{\psi_{\bm{\theta}}} = {U} (\bm{\theta}) \ket{\psi}$ is then detected via a set of measurements, represented by a positive operator-valued measure (POVM)~\cite{Helstrom:1976ij} $\{ \Pi_{x}\}$, where the POVM elements are labeled by an index $x$ that represents the possible outcomes (discrete or continuous) according to Born's rule $p(x |\bm{\theta}) = \bra{\psi_{\bm{\theta}}} \Pi_{x} \ket{\psi_{\bm{\theta}}}$. 

Often, the protocol is repeated $N$ times using identical copies of the state. From the output results $\mathbf{x} = (x_{1}, \ldots, x_{N})^{\top}${(the superscript $\top$ denotes the transpose)}, which we assume to be independent and identically distributed, one can construct a joint probability distribution $p(\mathbf{x}|\bm{\theta}) = \prod_{j=1}^{N} p(x_{j}|\bm{\theta})$ and then infer the vector parameter via an estimator $\widehat{\bm{\theta}}$.  In the following, we restrict our attention to locally unbiased estimators, for which $\langle \widehat{\theta}_{j} \rangle = \theta_{j}$ and $d\langle \widehat{\theta}_{j} \rangle/d\theta_{k} = \delta_{{jk}}$, and we compute average values $\langle \cdot \rangle$ using the probability distribution $p(\mathbf{x}|\bm{\theta})$~\cite{Helstrom:1976ij}.

The performance of the estimator can be properly assessed in terms of the covariance matrix $\mathbf{C}_{\psi} (\widehat{\bm{\theta}})$, defined as
\begin{equation}
[ \mathbf{C}_{\psi} (\widehat{\bm{\theta}})]_{jk} = 
\langle ( \widehat{\theta}_{j} - \theta_{j}) 
( \widehat{\theta}_{k} - \theta_{k} ) \rangle \, ,
\end{equation}
where $j, k = 1, \ldots, d$. The diagonal elements are the variances of the different parameters, whereas the nondiagonal elements characterize the possible correlations between various parameters.

% In a typical experiment,  it is not possible to calculate the exact values of the covariance matrix; we can only hope to set some lower limits on it. 
To guide the design of real experiments, it is possible to calculate theoretical lower bounds for $\mathbf{C}_{\psi} (\widehat{\bm{\theta}})$. The ultimate limit for any possible POVM is given by the time-honoured QCRB, which stipulates that
\begin{equation}
\mathbf{C}_\psi (\widehat{\bm{\theta}})  \succcurlyeq \mathbf{Q}_{\psi}^{-1}(\bm{\theta}) \, ,
\end{equation} 
where the matrix inequality $\mathbf{A} \succcurlyeq \mathbf{B}$ means that $\mathbf{A} - \mathbf{B}$ is a positive semidefinite matrix. Here, the quantum Fisher information matrix (QFIM) for pure states and unitary evolution takes the particularly simple form~\cite{Sidhu:2020aa}
\begin{equation}
\left[\mathbf{Q}_\psi (\bm{\theta}) \right]_{jk}= 4 \, \mathbf{C}_{\psi} (H_j,H_k) \, .
\end{equation}
The operators $H_{j}$ are the generators of the transformation, determined through $H_{j} =  iU^{\dagger} (\bm{\theta})\partial_{\theta_j} U(\bm{\theta})$, and we define the symmetrized covariance between two operators as $\mathbf{C}_{\psi} (A,B)=\frac{1}{2}\langle AB + BA\rangle - \langle A\rangle\langle B\rangle$ and expectation values with respect to $\ket{\psi}$. 

How can we optimize a probe state $\ket{\psi}$ given a \emph{matrix} figure of merit? Starting from a covariance matrix, we can balance the precision of the various parameters by using a weight matrix $\mathbf{W} \succ 0$; this leads to the scalar inequality
\begin{equation}
\textsc{wMSE} (\widehat{\bm{\theta}}) \equiv \Tr [ \mathbf{W} \mathbf{C}_{\psi} (\widehat{\bm{\theta}})] \ge \Tr [ \mathbf{W} \mathbf{Q}_{\psi}^{-1} (\bm{\theta}) ] \equiv C_{S}(\bm{\theta}) \, . 
\end{equation}
The left-hand side is the so-called weighted mean square error of the estimator $\textsc{wMSE}$, whereas $C_{S} (\bm{\theta})$ is a scalar cost function, much in the same spirit as the risk functions used in Bayesian estimation~\cite{Kay:1993aa}. For a given $\mathbf{W}$, the standard approach is to minimize $C_{S} (\bm{\theta})$ to derive better parameter estimates. In contradistinction, we address here whether there is an intrinsic choice for the matrix $\mathbf{W}$.

\emph{Intrinsic bound.---} 
Let us assume that the unitary process can be represented as $U(\bm{\theta}) = \exp [ i \bm{\Omega}_{(\bm{\theta})} \cdot \mathbf{X}]$, where the vector  $\bm{\Omega}_{(\bm{\theta})}$ encodes the pertinent set of parameters $\bm{\theta}$ and $\{ X_{i} \}$ are the generators of SU($n$); that is, $\{ X_{i} \}$ comprise a basis of the Lie algebra $\mathfrak{su}(n)$, which consists of traceless Hermitian $n\times n$ complex matrices with  $ i [ \cdot  , \cdot ]$ as their Lie bracket~\cite{Cornwell:1984aa} ($ [ \cdot, \cdot ]$ denoting the regular commutator). This form is especially advantageous because it includes all metrological applications in which the parameters are related to either components of a Hamiltonian or an interaction time. For our final results to hold, we require that the vector $\bm{\Omega_{(\theta)}}$ is a smooth injective function  of the parameters $\bm{\theta}$.

A straightforward method for calculating $\partial_{\theta_j} U(\bm{\theta})$ is through Wilcox's formula~\cite{Wilcox:1967aa} for a unitary operator $U= \exp(A)$:
\begin{equation}
\frac{\partial U}{\partial  \lambda} = \int_{0}^{1} \exp[(1- \beta)A] \, 
\frac{\partial A}{\partial \lambda} \, \exp( \beta A) \, d \beta \, .
\end{equation}
By recalling the definition of the generators, we find
\begin{align}
H_{j} = - \frac{\partial \bm{\Omega}_{\bm{\theta}}}{\partial \theta_j}\cdot \int_0^1 U^{-\beta}\, \mathbf{X} \, U^{\beta} d\beta \, .
\label{eq:Hj from integral}
\end{align} 
Since $U^{-1} \, \mathbf{X} \, U \in \mathfrak{su} (n)$, we conclude that $H_{j}$ can be generically expressed as a linear combination of the generators $\{X_i\}$ through
\begin{equation}
H_{j} =
\boldsymbol{\mathsf{h}}_j \cdot \mathbf{X} \, ,
\label{eq:generic generators}
\end{equation}
where the real vectors of coefficients $\boldsymbol{\mathsf{h}}_j(\bm{\theta})$ are computed from Eq.~\eqref{eq:Hj from integral} and depend on $\bm{\Omega}_{\bm{\theta}}$. The far-reaching result that the generators $H_j$ of the transformation are equal to linear combinations of the generators $\{X_i\}$ is essential to finding an intrinsic QCRB that is independent of parametrization. Moreover, this relation will always hold for the same vectors $\boldsymbol{\mathsf{h}}_j$ regardless of the particular representation of the group that we use.  

In this way, we can immediately work out a compact expression for the QFIM:
\begin{equation}
\label{eq:elQFIM}
\mathbf{Q}_{\psi} (\bm{\theta}) = 4 
\boldsymbol{\mathsf{H}}^{\top} (\bm{\theta}) 
\, \mathbf{C}_{\psi}(\mathbf{X}) \, \boldsymbol{\mathsf{H}}(\bm{\theta}) \, \end{equation} 
and the scalar QCRB now reads
\begin{equation}
\Tr [ \mathbf{W} \mathbf{C}_{\psi} (\widehat{\bm{\theta}})] \geq   \tfrac{1}{4} 
\Tr [\boldsymbol{\mathsf{H}}^{-1 \top}(\bm{\theta}) \mathbf{W} \boldsymbol{\mathsf{H}}^{-1}(\bm{\theta})
\, \mathbf{C}_{\psi}^{-1}(\mathbf{X}) ] \, .
\label{eq:QCRB with spit state and parameter dependence}
\end{equation}
The remarkable property of these expressions is that we have separated the parameter dependence $\boldsymbol{\mathsf{H}} (\bm{\theta})=\begin{pmatrix}\boldsymbol{\mathsf{h}}_1,& \ldots  &, \boldsymbol{\mathsf{h}}_{d} \end{pmatrix}^{\top}$ from the state dependence that is embodied in $[\mathbf{C}_{\psi} (\mathbf{X}) ]_{jk} = \mathbf{C}_{\psi} (X_{j}, X_{k})$.  This form allows us to easily identify when the QFIM becomes singular, which implies that all $d$ parameters cannot be simultaneously estimated for some probe state and parametrization. For example, when $\mathbf{C}_{\psi} (\mathbf{X})$ is singular, the probe state will never be useful for estimating all $d$ parameters, while, when $\boldsymbol{\mathsf{H}}$ is singular, the coordinate system defined by the parametrization is singular at that specific set of parameters $\bm{\theta}$ regardless of the probe state. As discussed in Ref.~\cite{Goldberg:2018aa}, singularities in one coordinate system can be alleviated for specific parameters by switching to a new coordinate system; here, we show how a proper choice of weight matrix $\mathbf{W}$ removes all ambiguities that can arise from the choice of coordinate system.

To proceed, we note that the parameters $\bm{\theta}$ are coordinates of the group manifold (the parameters $\bm{\Omega_{(\theta)}}$ form an alternative set of coordinates of the manifold), which is compact and simply connected~\cite{Cornwell:1984aa}. One can define therein a local metric tensor and through that covariant operations and invariant quantities. As $\mathfrak{su}(n)$ is compact and semisimple, the Killing form $B(X, Y) = 2n \, \Tr (X^{\dagger} Y ) $ is nondegenerate and defines a scalar product $(X, Y) = B (X,Y)$ that, in turn, induces a natural metric $ds^{2} = (dU, dU)$, called the Cartan metric. Actually,  by expressing the metric as 
\begin{equation}
ds^2=\sum_{jk}g_{jk}d\theta_j d\theta_k
\end{equation} 
and computing  $dU= -i U \sum_j H_j d\theta_j$, we find that the metric explicitly depends on the generators from Eq.~\eqref{eq:generic generators} through $g_{jk} = \Tr(H_j H_k)$. We can then use the orthonormality of the generators $(X_j,X_k)=\delta_{ij}$ 
to determine the components of the metric tensor to be (up to an inessential global constant)
\begin{equation}
g_{jk} =  \boldsymbol{\mathsf{h}}_{j} \cdot 
\boldsymbol{\mathsf{h}}_{k} \, ,  
\end{equation}
or, in vector form, $\mathbf{g} = \boldsymbol{\mathsf{H}}^{\top} \boldsymbol{\mathsf{H}}$. It follows that, if we choose the Cartan metric as the weight  matrix $\mathbf{W}=\mathbf{g}$, all of the parametrization dependence cancels from Eq.~\eqref{eq:QCRB with spit state and parameter dependence} and the QCRB becomes
\begin{equation}
\Tr [ \mathbf{g} \, \mathbf{C}_{\psi} (\widehat{\bm{\theta}}) ] \geq \tfrac{1}{4} 
\Tr [\mathbf{C}_{\psi}^{-1}(\mathbf{X})] \, .
\label{eq:QCRB with metric weight}
\end{equation} 
The only remaining ingredients are the covariances of the generators of $\mathfrak{su}(n)$ with respect to the original state, or any state along the unitary orbit, as $\Tr [\mathbf{C}_{\psi}^{-1}(\mathbf{X})]=\Tr [\mathbf{C}_{\psi^\prime}^{-1}(\mathbf{X})]$, with $\ket{\psi^\prime} = U \ket{\psi}$. This is exactly what one requires physically and seems quite elegant. It nicely complements the single-parameter scenario in which the QFI for unitary operations takes the same value anywhere along the unitary orbit, which previously presented peculiar challenges to the estimation of multiple parameters.

\emph{Discussion.---} 
To see how our new result works, let us first consider the simplest case of SU(2), which describes rotations, the type of unitary transformation being estimated in standard polarimetry and magnetometry, among other examples. A general element is characterized by the Euler angle parametrization $U(\Phi,\Theta,\Psi) = e^{- i \Phi J_z} e^{- i \Theta J_y} e^{- i \Psi J_z}$,  where we have used the standard angular momentum notation $\mathbf{J}$  for the generators. Derivatives with respect to the parameters lead us to 
\begin{equation}
    \begin{aligned}
    \boldsymbol{\mathsf{h}}_\Phi & =(-\sin \Theta \cos \Psi,\sin \Theta \sin \Psi,\cos \Theta)^\top \, , \\
    \boldsymbol{\mathsf{h}}_\Theta & =(\sin \Psi,\cos \Psi,0)^\top \,, \\
    \boldsymbol{\mathsf{h}}_\Psi & =(0,0,1)^\top \, ,
\end{aligned}
\end{equation} 
which makes the matrix $\boldsymbol{\mathsf{H}}$ singular when, for example, $\Theta=0$.
The metric for this parametrization is
\begin{equation}
    ds^2=d\Phi^2+2\cos\Theta d\Phi d\Psi+d\Theta^2 +d\Psi^2,
\end{equation}
which directly cancels the singularities in $\boldsymbol{\mathsf{H}}$ through $\boldsymbol{\mathsf{H}}^{-1\top}\mathbf{g}\boldsymbol{\mathsf{H}}^{-1}=\openone$, as promised.

It is straightforward to show that in this case $\mathbf{C}_{\psi} (\mathbf{J})$ is singular if and only if the probe state is an eigenstate of some angular momentum projection; that is, proportional to $U (\bm{\theta}) \ket{Jm}$. States with any definite angular momentum projection cannot be used for simultaneously estimating all three parameters of a rotation.

To find the most sensitive states we have to minimize $\Tr [ \mathbf{C}_{\psi}^{-1} (\mathbf{J})]$. This is straightforward to optimize because, for any symmetric, positive semidefinite, invertible matrix $\mathbf{M}$, $\Tr (\openone)^2 = \Tr ( \mathbf{M}^{1/2} \mathbf{M}^{-1/2})^2 \leq \Tr (\mathbf{M}) \Tr (\mathbf{M}^{-1})$, with equality if and only if $\mathbf{M}$ is proportional to the identity matrix. Since  $\Tr [ \mathbf{C}_{\psi} (\mathbf{J}) ] = J(J+1) - \sum_{k}\langle J_k\rangle^2 $, we find
\begin{equation}
    \Tr [\mathbf{C}_{\psi}^{-1} (\mathbf{J}) ] \geq \frac{9 }{J(J+1)} \, ,
    \label{eq:CRB for SU(2)}
\end{equation} 
with the trace of the inverse achieving the minimum only when the state is first-order unpolarized~\cite{Goldberg:2021tx} (that is, $\langle 	\mathbf{J} \rangle = 0$), and when $\mathbf{C}_{\psi}(\mathbf{J}) \propto \openone $, as is the case for the so called Kings of Quantumness~\cite{Bjork:2015aa,Bouchard:2017aa,Goldberg:2020aa}. We see that having isotropic angular momentum up until second order makes a state most sensitive to arbitrary rotations about arbitrary axes. Our minimum intrinsic QCRB is given by the square of the number of parameters divided by the quadratic Casimir invariant $\mathcal{C}_2= J_x^2 +J_y^2 + J_z^2 =J(J+1) \openone$.

The conditions guaranteeing the saturation of the QCRB constitute a touchy business~\cite{Fujiwara:1994aa}. Fortunately, for the case of pure states,  a sufficient condition is $\langle  [H_{j}, H_{k}] \rangle = 0,\, \forall j, k$.  These expectation values will vanish for all states with $\langle  \mathbf{J} \rangle = 0$ so the optimal state will have a saturable QCRB  in this case.

These results can be directly extended to the case of a full SU($n$) estimation. A large number of applications across photonics require the capability of implementing arbitrary linear unitary transformations on a set of optical modes. The control and estimation of all of the free parameters of a general modal unitary transformation is thus of paramount importance for many experiments. This includes spatial multiplexing~\cite{Morizur:2010ua,Bozinovic:2013um}, neural networks~\cite{Shen:2017vs}, novel imaging techniques~\cite{Popoff:2010to}, quantum circuits~\cite{Politi:2008tl}, and programmable simulators~\cite{Harris:2017uv}. Sampling the output probability when photons traverse such networks is hard to simulate in a classical computer and hence it may be a path to achieve quantum supremacy~\cite{Zhong:2020tj}.

Using the same trick, we get  that  $\Tr [ \mathbf{C}_{\psi} (\mathbf{X}) ]$ is bounded by the quadratic Casimir invariant $\mathcal{C}_2 = \sum_{jk} g_{jk} X_{j} X_{k}\equiv\tilde{\mathcal{C}}_2\openone$ of SU($n$) and, therefore, the second-order unpolarized  states
\begin{equation}
\langle  \mathbf{X} \rangle = 0 \, ,  \qquad 
\mathbf{C}_{\psi} (\mathbf{X} )=\frac{\mathcal{C}_2}{d} \,, 
\label{eq:second order unpolarized general}
\end{equation}
saturate the optimal QCRB for the simultaneous estimation of all $d$ {intrinsically weighted} parameters. This yields a minimum \textit{total} variance {$\tfrac{1}{4} d^2 \tilde{\mathcal{C}}_2$} {$d^2/4\tilde{\mathcal{C}}_2$}, generalizing Eq.~\eqref{eq:CRB for SU(2)}. All of the parameters should be simultaneously estimable with Heisenberg-scaling precisions, because the square root of the Casimir invariant $\sqrt{\tilde{\mathcal{C}}_2}$, which corresponds to physical quantities such as the number of particles, is quadratic in the labels of the irreducible representation.

Such a protocol outperforms strategies where the parameters are being sequentially estimated with uncorrelated measurements~\cite{Giovannetti:2006ud,Giovannetti:2011aa,Yuan:2016ta} by a factor on the order of $d$. Uncorrelated sequential estimation protocols require splitting the physical resources $\mathcal{O}(\sqrt{\tilde{\mathcal{C}}_2})$ into $d$ parts, then at best estimating \textit{each} of the $d$ parameters with variance $\mathcal{O}[({d}/{\sqrt{\tilde{\mathcal{C}}_2}})^{2} ]$, so that the \textit{total} variance scales as $\mathcal{O}(d^3 / \tilde{\mathcal{C}}_2)$. This has been seen for the specific examples of multiphase estimation~\cite{Humphreys:2013}, which satisfies U(1)$^{\otimes n} \subset $SU($n$) up to a global phase (i.e., modulo the centre of the group), and rotation estimation~\cite{Baumgratz:2016aa}, which satisfies SU(2)$ \subset $SU($n$). Estimating the components of any subset of SU($n$) can now be similarly optimized using our results.

\begin{figure}
    \centering
    \includegraphics[width=\columnwidth]{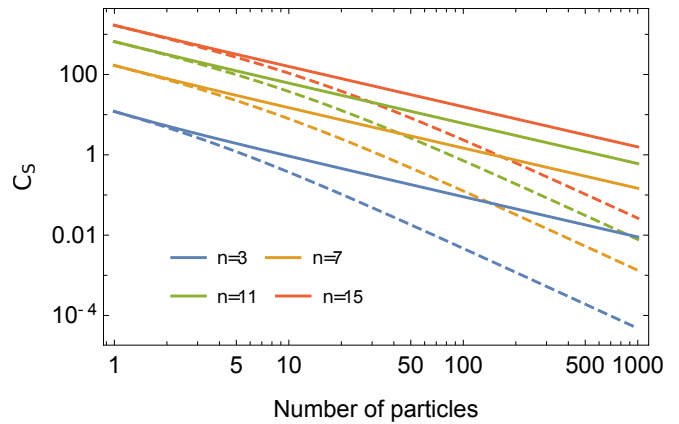}
    \caption{
    Minimum total uncertainty $C_S$ for estimating all of the parameters of an SU($n$) transformation for our ideal states satisfying Eq. \eqref{eq:second order unpolarized general} (dashed curves) and GHZ states (solid curves) versus number of particles (photons, qudits, etc.), using our intrinsic weighting of the parameters. While GHZ states, like NOON states, exhibit Heisenberg-scaling precision for estimating some parameters, they convey overall minimum uncertainties exhibiting shot-noise-scaling precision. The ideal states triumph with Heisenberg-scaling precision for the simultaneous estimation of all of the parameters.} 
    \label{fig:comparing GHZ to unpol}
\end{figure}

When estimating the $d=n^2-1$ free parameters of SU($n$) transformations among qudits, states satisfying Eq. \eqref{eq:second order unpolarized general} are again optimal and achieve the Heisenberg limit for simultaneously estimating all $d$ parameters. The relative advantage of such states over Greenberger-Horne-Zeilinger (GHZ) states~\cite{Greenberger:1990vf},  which have been deemed as of optimal sensitivity~\cite{Giovannetti:2004wc}, is depicted in Fig.~\ref{fig:comparing GHZ to unpol} for various values of $n$; GHZ states only achieve shot-noise scaling in this context. 
{This is because GHZ states are optimized for estimating only a subset of the parameters of an SU($n$) transformation, while the intrinsic weight matrix includes all of the parameters.}

The tools developed here are expeditious to apply. Consider, for example, SU(3) metrology, which governs physical systems such as symmetric states of $\mathcal{N}$ qutrits. SU(3) transformations are parametrized by eight variables, making the choice of weight matrix $\mathbf{W}$ cumbersome and potentially arbitrary: fixing $\mathbf{W}$ to correspond to SU($3$)'s metric dramatically simplifies the problem, yielding Eq. \eqref{eq:QCRB with metric weight}.

One immediate application is to polarimetry: while standard polarimetry estimates the polarization components of a transverse electric field~\cite{Goldberg:2020vj}, a general beam may have 3D polarization structure,  which requires the knowledge of the 3D Stokes vectors~\cite{Dennis:2004wh,Luis:2005ue,Gil:2010ub,Sheppard:2012tg}. The latter's eight degrees of freedom are exactly the parameters imprinted by SU(3) operations and constitute the basis of 3D polarimetry, which has plenty of applications in different areas. Our result uniquely simplifies the choice of relative weights for the variances and covariances between all eight parameters.

Equation \eqref{eq:second order unpolarized general} can readily be used to find ideal states for estimating all eight parameters in three-dimensional polarimetry{, without any qualms about how to choose the relative weights of these eight parameters}. The former leads to conditions among the coefficients of the states in the photon-number basis, among whose solutions is the set of states
\begin{equation}
    \ket{\psi}= \tfrac{1}{\sqrt{3}} ( \ket{k-\ell,k,k+\ell} + 
    \ket{k,k+\ell,k-\ell}+\ket{k+\ell,k-\ell,k} ) ,
\end{equation}
for pairs of nonzero integers $k=\mathcal{N}/3$ and $\ell$ satisfying $4\ell^2=3k(k+1)$. Here, the state $\ket{n_{1},n_{2}, n_{3}}$ denotes a Fock state with $n_{1}$, $n_{2}$, and $n_{3}$ excitations in three respective bosonic modes or a symmetric superposition of $n_{1}$, $n_{2}$, and $n_{3}$ copies of three respective states.  The relative merit of using such states versus GHZ states $(\ket{\mathcal{N},0,0} + \ket{0,\mathcal{N},0} +\ket{0,0,\mathcal{N}})/\sqrt{3}$ can be appreciated again in Fig.~\ref{fig:comparing GHZ to unpol}.

%%%%%%%%%%%%%%%
%%%%%%%%%%%%%

Two final comments pertain. First, it is easy to realize that the elegant form (\ref{eq:elQFIM}) for the QFIM also holds for mixed states, where we must employ the replacement~\cite{Petz:2011aa}
\begin{equation}
    \mathbf{C}_{\psi}(\mathbf{X}) \mapsto 
\frac{1}{2}\int_0^\infty ds\Tr\left[\left(\varrho \mathbf{X}-\mathbf{X}\varrho\right)e^{-\varrho s}\left(\mathbf{X}^\top\varrho-\varrho \mathbf{X}^\top\right)e^{-\varrho s}\right],
\label{eq:replacement for mixed}
\end{equation}
so the parameter dependencies factor out in terms of $\mathbf{H}(\bm{\theta})$ and  will always disappear in the QCRB if we weight the covariances using the metric. This means that even noisy measurements with imperfect probe states can benefit from our prescription to choose the weight matrix $\mathbf{W}$ to coincide with the metric describing the transformation by using Eq.~\eqref{eq:QCRB with metric weight} with the replacement of Eq.~\eqref{eq:replacement for mixed}. Second, the treatment can be repeated much in the same way when the information is encoded either in any of the subgroups of SU($n$) or in channels with symmetries beyond SU($n$). {This lets the results take into account scenarios in which not all of the parameters are physically relevant, such as recovering the optimality of NOON states when estimating a single relative phase as a subset of SU($2$).}

\emph{Concluding remarks.---}
We have used geometric arguments to show how to obtain an intrinsic QCRB for unitary processes. This directly generalizes many earlier results for multiparameter estimation, notably generalizing the single-parameter estimation result that estimation precision is unchanged along a unitary orbit, and is easily applicable to variegated metrological scenarios. An important advantage of these geometric arguments is that they provide deep connections between quantum estimation and general quantum information tasks.   

\emph{Acknowledgment.---}
We would like to thank Hubert de Guise and Pieter Kok for useful discussions. AZG acknowledges funding from NSERC, the Walter C. Sumner Foundation and Cray Inc. LLSS acknowledges financial support from the European Union’s Horizon 2020 research and innovation program (Projects ApresSF and Stormytune) and the Spanish Ministerio de Ciencia e Innovación (Grant No. PGC2018- 099183-B-I00). HF acknowledges funding from NSERC and CIFAR.

%\bibliography{metrology}

\end{document}